\documentclass{article}
\usepackage{spconf,amsmath,graphicx}
\usepackage{comment,psfrag,amssymb,booktabs,tabularx} 

\title{A Hybrid Motion Estimation Technique for Fisheye Video Sequences Based on Equisolid Re-Projection}
\name{Andrea Eichenseer, Michel B\"atz, J\"urgen Seiler, and Andr\'e Kaup}
\address{Multimedia Communications and Signal Processing\\
	Friedrich-Alexander University Erlangen-N\"urnberg (FAU), Cauerstr. 7, 91058 Erlangen, Germany\\
	}
\begin{document}
%
\maketitle
\begin{abstract}

Capturing large fields of view with only one camera is an important aspect in surveillance and automotive applications, but the wide-angle fisheye imagery thus obtained exhibits very special characteristics that may not be very well suited for typical image and video processing methods such as motion estimation. This paper introduces a motion estimation method that adapts to the typical radial characteristics of fisheye video sequences by making use of an equisolid re-projection after moving part of the motion vector search into the perspective domain via a corresponding back-projection. By combining this approach with conventional translational motion estimation and compensation, average gains in luminance PSNR of up to 1.14 dB are achieved for synthetic fisheye sequences and up to 0.96 dB for real-world data. Maximum gains for selected frame pairs amount to 2.40 dB and 1.39 dB for synthetic and real-world data, respectively.

\end{abstract}
\section{Introduction}
\label{sec:intro}

\begin{figure}[t]
\centering
\centerline{\includegraphics[width=\columnwidth]{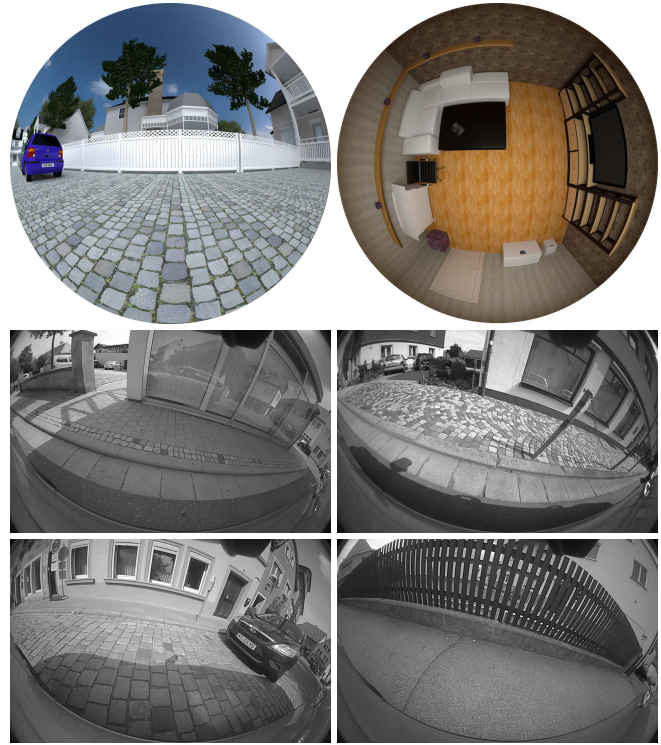}}
\caption{Example frames of synthetic (top) and real-world fisheye data (bottom). Top left to bottom right: \textit{Street}, \textit{Room}, \textit{Video1}, \textit{Video2}, \textit{Video3}, and \textit{Video4}.}
\label{fig:6exmp}
\vspace{-0.3cm}
\end{figure}

In video surveillance as well as in automotive applications, it is often necessary to capture very large fields of view (FOV) well beyond the common 50 or 60 degrees.
Fisheye lenses~\cite{miyamoto1964fel} are capable of achieving an FOV of 180 degrees and more, thereby introducing strong radial distortion into an image.
Fig.~\ref{fig:6exmp} shows examples of synthetic circular as well as real-world full-frame fisheye images.

Although fisheye images contain the scene information of an entire hemisphere, the resulting characteristics may lead to drawbacks when performing traditional image and video processing methods such as motion estimation (ME).
As the predominant kind of motion in a video is assumed to be translation, typical block-matching ME techniques~\cite{blockmatching} are based on translational models.
This kind of ME is also adopted in hybrid video coding standards such as the widely used H.264/Advanced Video Coding (AVC)~\cite{wiegand2003avc} as well as its successor,  H.265/High Efficiency Video Coding (HEVC)~\cite{sullivan2012hevc}.
Application scenarios are, naturally, not limited to video coding.
Motion estimation is strongly related to optical flow estimation~\cite{tvl1} and forms an integral part in error concealment~\cite{dmve}, image registration and super resolution~\cite{park2003super}, as well as frame rate up-conversion~\cite{sunwoo}, to name only a few more application scenarios.
While many established ME methods work well for conventional rectilinear image and video content, they are certainly not designed for non-translational motion content.
In such cases, adaptations of existing methods may prove very beneficial, as shown in~\cite{narroschke2013affine}, where an affine transformation model is added to the ME procedure of HEVC.
Especially in the field of computer vision, fisheye lenses and omnidirectional cameras are a common topic of interest and there exist numerous publications on new camera models and algorithms adapted for fisheye imagery~\cite{kannala2006genericmodel, compvis2, compvis}.

In a similar fashion, we introduce a hybrid approach for block-based motion estimation that takes into account the special fisheye characteristics.
The simulation results obtained in~\cite{eichenseer2014dcorfish} indicate that for fisheye sequences, techniques such as motion estimation might be better conducted in the distortion-corrected, i.\,e., the perspective domain.
Our approach further investigates and substantiates this assumption.

The remainder of the paper is structured as follows.
Section~\ref{sec:stateoftheart} provides a brief summary of block-based translational motion estimation and compensation. In Section~\ref{sec:equisolid}, we present our motion estimation approach for fisheye sequences based on equisolid re-projection and implement it as a hybrid method in Section~\ref{sec:hybrid}. Section~\ref{sec:results} provides simulation results and Section~\ref{sec:conclusion} concludes this paper.

\section{Motion Estimation based on a Translational Model}
\label{sec:stateoftheart}

Block-based translational motion estimation is a widely employed technique based on matching blocks from the frame to be predicted 
to a reference frame,
which is typically a temporal neighbor of this frame.
For a pre-defined search range of pixels or even sub-pixels, motion vector candidates are evaluated for each block 
of a given block size using an error metric such as the sum of absolute differences (SAD) or the sum of squared differences (SSD).
The motion vector candidate 
that leads to a minimization of the residual error 
is saved for the current block
and then used for copying the motion compensated block from the reference frame. 
Doing this for each block of the frame leads to a motion compensated frame that can then be used as a predictor.

While this approach works very well for videos with translational motion, there is a drawback when dealing with radially distorted imagery such as fisheye videos.
Since fisheye images do not follow the projective geometry principle of mapping a straight line again onto a straight line for translational motion, a motion compensation method based on this kind of model cannot yield optimum results.

\section{Motion Estimation via Equisolid Re-Projection}
\label{sec:equisolid}

\begin{figure}[t]
\centering
\centerline{\includegraphics[width=\columnwidth]{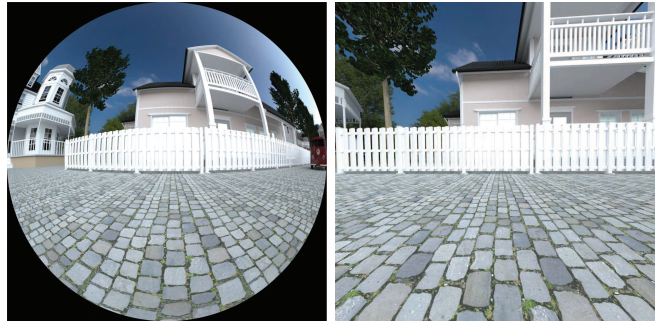}}
\vspace{-0.2cm}
\caption{Comparison of an equisolid fisheye frame (left) to its perspective version (right) assuming the same sensor size.}
\label{fig:rectifiedexmp}
\vspace{-0.3cm}
\end{figure}

For perspective projection, the pinhole model holds.
In this model, the incident ray of light $\theta$, whose maximum value corresponds to half the FOV, is projected onto the image plane at a distance
\vspace{-0.2cm}
\begin{equation}
r_\mathrm{p} = f \tan\theta
\vspace{-0.1cm}
\end{equation}
to the image center, where $f$ is the focal length.
Clearly, the tangent function is a limiting factor here, as it does not allow for incident rays of light with an angle of more than 90 degrees, i.\,e., the FOV is limited to just below 180 degrees.
Although mathematically possible, such an FOV would result in near-infinite sensor dimensions.
Fig.~\ref{fig:rectifiedexmp} shows the limits with regard to FOV for perspective images.
Consequently, the pinhole model is not applicable to very wide FOVs.

In order to capture a very wide viewing angle of up to and beyond 180 degrees, another projection function must be chosen.
One such projection is equisolid angle projection~\cite{miyamoto1964fel,kannala2006genericmodel}, which is expressed as
\vspace{-0.2cm}
\begin{equation}
\label{eq:equisolid}
r_\mathrm{e} = 2 f \sin(\theta/2)\:.
\vspace{-0.1cm}
\end{equation}
In contrast to the pinhole model, the equisolid model can map a much larger FOV onto a sensor of much smaller size.
For fisheye data that is based on this equisolid model, we now design a motion estimation technique in a fashion similar to~\cite{eichenseer2014dcorfish}, where the fisheye data is distortion corrected prior to coding so as to save bit rate.





Since we want to perform motion estimation on fisheye frames, we cannot remove the radial distortion and then use the perspective frames for further processing.
Instead, we implement the idea of switching to the perspective domain in a different manner.
An equisolid-to-perspective projection can be expressed as
\vspace{-0.2cm}
\begin{equation}
\label{eq:backward}
r_\mathrm{p} = f \tan\left(2\arcsin\left(\frac{r_\mathrm{e}}{2f}\right)\right)\:.
\vspace{-0.2cm}
\end{equation}
Performing a re-projection back to the equisolid domain is then defined as
\vspace{-0.2cm}
\begin{equation}
\label{eq:forward}
r_\mathrm{e} = 2 f \sin\left(\frac{1}{2} \arctan\left(\frac{r_\mathrm{p}}{f} \right) \right)\:.
\vspace{-0.2cm}
\end{equation}

As motion estimation based on block-matching is assumed to yield better results for perspective imagery due to the translational model employed, switching to this domain seems to be a reasonable step.
However, actually transforming the image into the perspective domain would be highly impractical due to the vast amount of pixels this would require.
Hence, we instead manipulate the pixel coordinates in a suitable fashion as described in the following.

For each block of the current fisheye frame, the pixel positions are stored and projected into the perspective domain using~(\ref{eq:backward}).
The motion vector candidate, which is taken from within a defined integer-pixel search range, is then added to these pixel positions, effectively performing a translational motion.
These translated pixel positions are then re-projected into the equisolid fisheye domain using~(\ref{eq:forward}).

Having obtained the manipulated pixel positions, the corresponding pixel values can be copied from the reference frame.
The reference frame is upsampled beforehand using a suitable interpolation technique.
This is necessary since the new pixel positions are no longer located at integer positions but somewhere in between the integer pixel grid.
These pixel values then yield a compensated block candidate that must be evaluated according to an error minimization criterion, as in translational motion estimation.
Performing this technique for all blocks of the frame then yields the compensated frame.

\begin{figure}[t]
\centering
\psfrag{S}[cc][cB][1.00][0]{SSD}
\psfrag{o}[cc][cB][1.00][0]{$<$}
\psfrag{b}[cc][cc][1.00][0]{translational ME}
\psfrag{P}[cc][Bc][1.25][0]{$\mathcal{P}$}
\psfrag{E}[cc][Bc][1.25][0]{$\mathcal{E}$}
\psfrag{M}[cc][cc][1.0][0]{motion vector candidate}
\centerline{\includegraphics[width=\columnwidth]{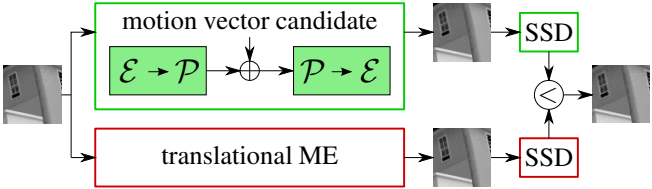}}
\vspace{-0.2cm}
\caption{Schematic depiction of the hybrid ME technique for one block. Top path: equisolid ME showing the two projections performed on the pixel positions of the block, bottom path: traditional block-based translational ME.}
\label{fig:blockdiagramtex}
\vspace{-0.3cm}
\end{figure}

\section{Hybrid Motion Estimation for Fisheye Sequences}
\label{sec:hybrid}

The technique described above effectively reduces any finite search range according to~(\ref{eq:forward}), thus making it impossible to correctly detect motion in the periphery of the fisheye image.
We hence propose a hybrid approach, combining the equisolid projection-based ME with translational ME.
As a block-based decision criterion, we choose the sum of squared differences.
For each block, the method that yields the lower SSD value is chosen as the motion compensation method.

This procedure is schematically depicted in Fig.~\ref{fig:blockdiagramtex}, where $\mathcal{E}$ and $\mathcal{P}$ denote what, in this paper, we call the equisolid and the perspective domain, respectively.
The top path describes the equisolid re-projection approach with the motion vector addition in the perspective domain, while the lower one represents traditional block-based ME.
The two compensated block candidates obtained are compared with regard to the SSD and the better candidate is chosen as the final compensated block.
The following section analyzes this hybrid approach for synthetically generated as well as real-world data.

\begin{figure}[t]
\centering
\centerline{\includegraphics[width=\columnwidth]{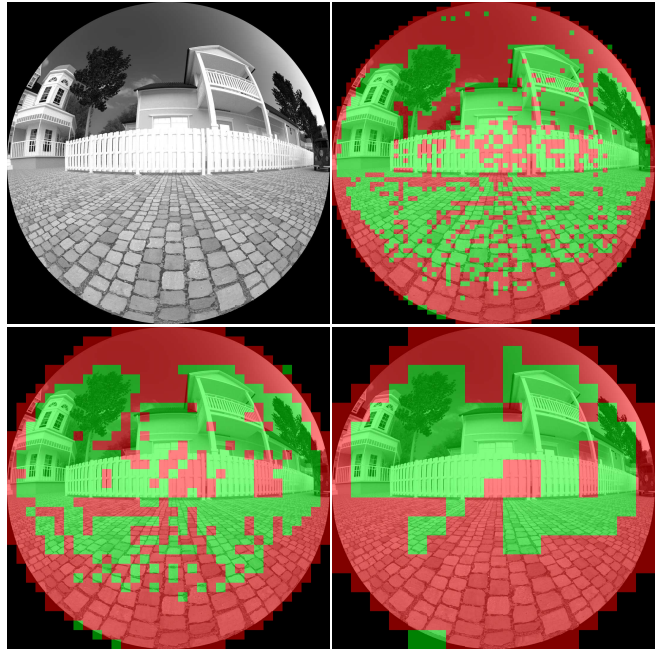}}
\vspace{-0.2cm}
\caption{Hybrid ME for frames 1500 and 1501 of \textit{Street}. Top: compensated frame and overlaid decision map for block size 16$\times$16. Bottom: Decision maps for block sizes 32$\times$32 and 64$\times$64. Red: translational ME, green: equisolid ME.}
\label{fig:cars1500masks}
\vspace{-0.4cm}
\end{figure}

\begin{figure}[t]
\centering
\centerline{\includegraphics[width=\columnwidth]{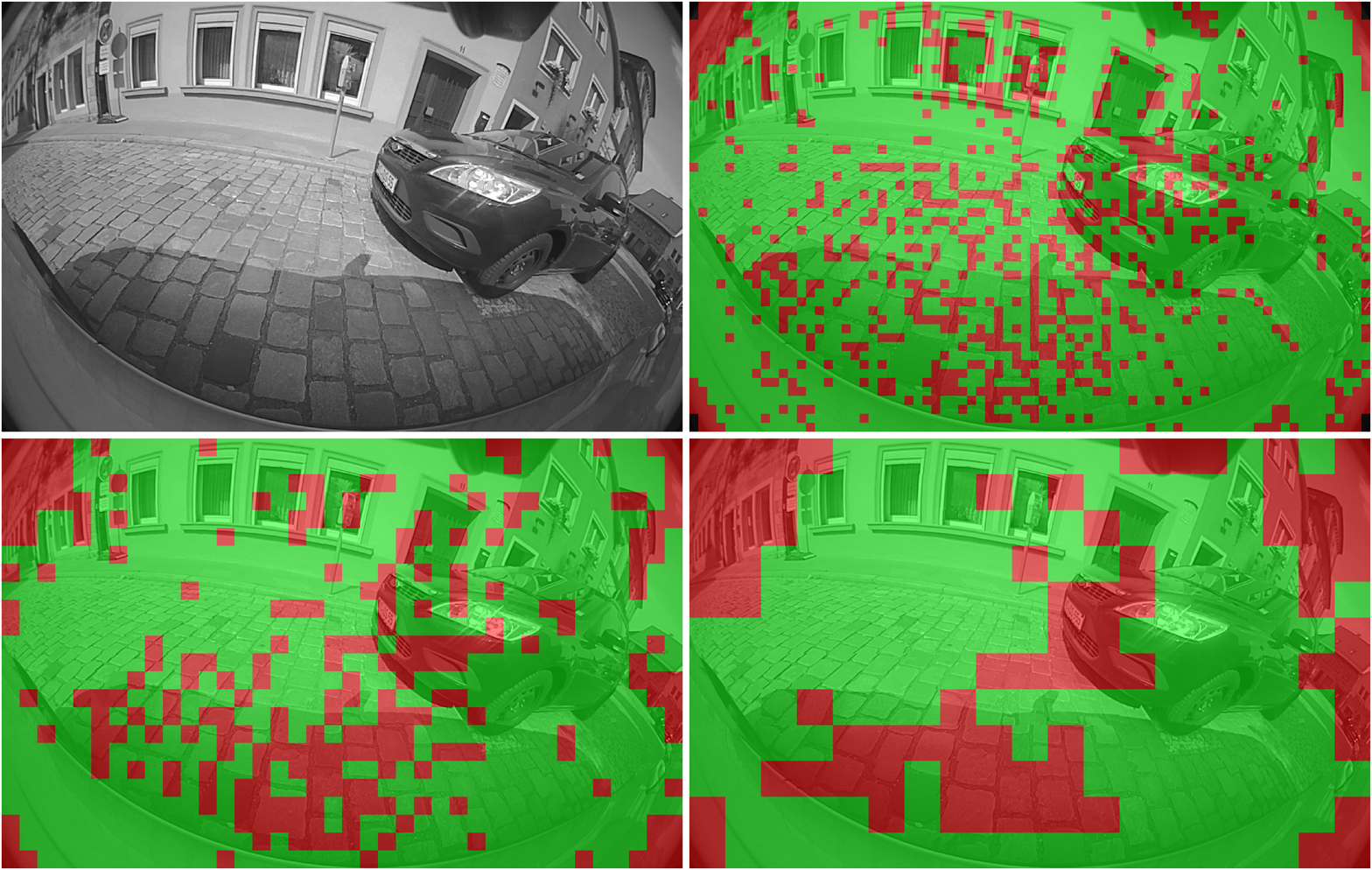}}
\vspace{-0.2cm}
\caption{Hybrid ME for frames 5 and 6 of \textit{Video3}. Top: compensated frame and overlaid decision map for block size 16$\times$16. Bottom: Decision maps for block sizes 32$\times$32 and 64$\times$64. Red: translational ME, green: equisolid ME.}
\label{fig:video3_5masks}
\vspace{-0.4cm}
\end{figure}

\section{Simulation Setup and Results}
\label{sec:results}

\begin{table*}[t]
\caption{Average luminance PSNR results in dB for a search range of 128 pixels and varying block size.}
\label{tab:psnr}
\vspace{0.1cm}
\centering
\renewcommand\arraystretch{0.9}
\begin{tabularx}{\textwidth}{p{1cm}p{0cm}cccp{0cm}cccp{0cm}cccp{0cm}cccp{0cm}c}
\toprule
& & \multicolumn{3}{c}{Block size 8$\times$8} & & \multicolumn{3}{c}{Block size 16$\times$16} & & \multicolumn{3}{c}{Block size 32$\times$32} & &  \multicolumn{3}{c}{Block size 64$\times$64} & & \\
 & & TME & HME & $\Delta$ & & TME & HME & $\Delta$ & & TME & HME & $\Delta$ & & TME & HME & $\Delta$ & & \\
\midrule
\textit{Street} & & 32.44 & 34.12 & \textbf{1.68} & & 30.95 & 32.32 & \textbf{1.37} & & 29.03 & 30.06 & \textbf{1.03} & & 26.53 & 27.22 & \textbf{0.68} & & \\
\textit{Room} & & 46.89 & 47.51 & \textbf{0.61} & & 46.24 & 46.77 & \textbf{0.53} & & 45.14 & 45.55 & \textbf{0.41} & & 43.94 & 44.40 & \textbf{0.45} & & \\
\addlinespace
Average & & & & 1.14 &  &  &  & 0.95 &  & & & 0.72 & &  &  & 0.57 & & \\
\midrule
\textit{Video1} & & 35.69 & 36.82 & \textbf{1.13} & & 33.47 & 34.36 & \textbf{0.89} & & 31.28 & 32.06 & \textbf{0.78} & & 28.81 & 29.64 & \textbf{0.84} & & \\
\textit{Video2} & & 31.89 & 32.73 & \textbf{0.84} & & 28.64 & 29.41 & \textbf{0.77} & & 25.50 & 26.24 & \textbf{0.74} & & 23.03 & 23.68 & \textbf{0.65} & & \\
\textit{Video3} & & 33.76 & 34.81 & \textbf{1.05} & & 31.42 & 32.28 & \textbf{0.86} & & 28.86 & 29.58 & \textbf{0.71} & & 26.24 & 26.92 & \textbf{0.68} & & \\
\textit{Video4} & & 33.25 & 34.07 & \textbf{0.82} & & 30.89 & 31.69 & \textbf{0.80} & & 28.28 & 29.15 & \textbf{0.87} & & 25.63 & 26.49 & \textbf{0.87} & & \\
\addlinespace
Average & & & & 0.96 &  &  & & 0.83  & &  &  & 0.77 & &  &  & 0.76 & & \\
\bottomrule
\end{tabularx}
\vspace{-0.4cm}
\end{table*}


For testing our approach, we created synthetic equisolid fisheye data in blender~\cite{blender}, combining several object models from~\cite{blendswap} to create suitable scenes.
The camera was set to a panoramic fisheye (equisolid projection) with an FOV of 185 degrees, a focal length of 1.8 mm, and a sensor size of 5.2 mm square that is just able to capture the entire circular fisheye.
Example frames of our synthetic videos are given in the top row of Fig.~\ref{fig:6exmp}.
\textit{Room} uses a static camera and various moving objects, while \textit{Street} uses a moving camera and static objects and is thus purely translational in motion.
Both have a resolution of 1088$\times$1088 pixels at 25 frames per second.

To additionally test our approach on real-world data, we used traffic video sequences with global translational motion.
Example frames taken from the middle of each sequence are shown in rows two and three of Fig.~\ref{fig:6exmp}.
We assume the same FOV and focal length as used in the synthetic images, but since the real-world data actually represents full-frame fisheye videos, we derive a sensor size of 4.6 mm by 2.9 mm by looking for the maximum radius that could be mapped onto the image plane.
This derivation is based on the assumption that 5.2 mm is able to represent the circular fisheye.
The real-world sequences are also purely translational in motion and correspond to aligned versions of the ones used in~\cite{eichenseer2014dcorfish}.
Their resolution is 1216$\times$768 pixels at 15 frames per second for \textit{Video2}, and 30 frames per second for the other three videos.

In our implementation, we used a fixed search range of 128 pixels in every direction for both the translational ME as well as the equisolid ME so that in each method, 66049 integer-pixel motion vector candidates were evaluated for each block.
In the equisolid ME, the re-projected pixel positions were then applied to a reference frame that was upscaled by a factor of 8 via cubic convolution interpolation.
The proposed hybrid approach was tested for block sizes of 8 to 64 pixels square and evaluated for 10 frame pairs of \textit{Street} (placed at 100 frame intervals), 3 frame pairs of \textit{Room} (placed at 150 frame intervals), and 30 consecutive frames each of \textit{Video1} through \textit{Video4}.

Fig.~\ref{fig:cars1500masks} shows a prime example of a compensation result from the hybrid method along with overlaid decision maps for different block sizes.
Green blocks used the equisolid ME, red blocks the translational ME.
At the periphery of the circular fisheye, the search range limitation of equisolid ME becomes quite evident. 
Fig.~\ref{fig:video3_5masks} shows a corresponding example for the real-world fisheye sequences.
These examples are representative for all frame pairs tested.

Average luminance PSNR results comparing translational motion estimation (TME) only and the proposed hybrid motion estimation (HME) technique are summarized in Table~\ref{tab:psnr}, where the block size used is given in pixels.
From these results, it is evident that our proposed hybrid method is able to improve on translational ME for all sequences tested.
Although only the synthetic sequences exactly follow~(\ref{eq:equisolid}) and while it can be expected that real fisheye lenses never project incoming light rays according to an exactly equisolid projection function, it is clearly a justified assumption.
As the real-world sequences are not circular, thus effectively cutting off the hard to predict periphery, the equisolid approach covers even more blocks. 
Following from~(\ref{eq:forward}), the search range decreases significantly for content lying closer to the image boundaries and it thus represents, along with the tangent function, a limiting factor of the equisolid ME technique.
Accordingly, we would expect higher gains for higher search ranges.

\section{Conclusion}
\label{sec:conclusion}

This paper introduced a motion estimation method specifically designed for equisolid fisheye video sequences.
Furthermore, a hybrid technique combining this equisolid ME with translational ME was proposed as a practical implementation.
The results showed that employing the proposed hybrid ME technique achieves an average gain of up to 1.14 dB in luminance PSNR compared to using translational motion estimation only.
For selected frame pairs, even higher gains of up to 2.40 dB were obtained.
It was also shown that the proposed method works well not only for synthetically generated material but also for real-world data captured with actual fisheye cameras. Here, an average gain of up to 0.96 dB was achieved. Maximum gains for selected frame pairs amounted to 1.39 dB.

Tackling the problem of overstepping the 180 degree boundary as well as examining the limitations with regard to the search range are part of current research.
Other points of investigation are the influence of sub-pixel accurate motion vector candidates in the perspective domain, the influence of the employed interpolation of the reference frame, and the use of polynomial approximations for the projection functions.
Future work will furthermore explore our method's suitability for video coding.

\section{Acknowledgment}
\vspace{-0.15cm}

This work was partly supported by the Research Training Group 1773 “Heterogeneous Image Systems”, funded by the German Research Foundation (DFG).
The real-world fisheye sequences were kindly provided by Continental Chassis \& Safety  BU ADAS Segment Surround View (A.D.C GmbH), Kronach.


\bibliographystyle{IEEEbib}
\bibliography{refs}

\begin{thebibliography}{10}

\bibitem{miyamoto1964fel}
K.~Miyamoto,
\newblock ``{Fish Eye Lens},''
\newblock {\em Journal of the Optical Society of America}, vol. 54, no. 8, pp.
  1060--1061, August 1964.

\bibitem{blockmatching}
M.~Santamaria and M.~Trujillo,
\newblock ``{A Comparison of Block-Matching Motion Estimation Algorithms},''
\newblock in {\em Proceedings of the IEEE 7th Colombian Computing Congress},
  October 2012, pp. 1--6.

\bibitem{wiegand2003avc}
T.~Wiegand, G.~J. Sullivan, G.~Bjontegaard, and A.~Luthra,
\newblock ``{Overview of the H.264/AVC Video Coding Standard},''
\newblock {\em IEEE Transactions on Circuits and Systems for Video Technology},
  vol. 13, no. 7, pp. 560--576, July 2003.

\bibitem{sullivan2012hevc}
G.~J. Sullivan, J.~Ohm, W.-J. Han, and T.~Wiegand,
\newblock ``{Overview of the High Efficiency Video Coding (HEVC) Standard},''
\newblock {\em IEEE Transactions on Circuits and Systems for Video Technology},
  vol. 22, no. 12, pp. 1649--1668, December 2012.

\bibitem{tvl1}
J.~Sanchez, E.~Meinhardt-Llopis, and G.~Facciolo,
\newblock ``{TV-L1 Optical Flow Estimation},''
\newblock {\em {Image Processing On Line}}, vol. 3, pp. 137--150, July 2013.

\bibitem{dmve}
J.~Zhang, J.~F. Arnold, and M.~R. Frater,
\newblock ``{A Cell-Loss Concealment Technique for MPEG-2 Coded Video},''
\newblock {\em IEEE Transactions on Circuits and Systems for Video Technology},
  vol. 10, no. 4, pp. 659--665, June 2000.

\bibitem{park2003super}
S.~C. Park, M.~K. Park, and M.~G. Kang,
\newblock ``{Super-\-Resolution Image Reconstruction: A Technical Overview},''
\newblock {\em {IEEE Signal Processing Magazine}}, vol. 20, no. 3, pp. 21--36,
  May 2003.

\bibitem{sunwoo}
U.~S. Kim and M.~H. Sunwoo,
\newblock ``{New Frame Rate Up-Conversion Algorithms With Low Computational
  Complexity},''
\newblock {\em IEEE Transactions on Circuits and Systems for Video Technology},
  vol. 24, no. 3, pp. 384--393, March 2014.

\bibitem{narroschke2013affine}
M.~Narroschke and R.~Swoboda,
\newblock ``{Extending HEVC by an Affine Motion Model},''
\newblock in {\em Proceedings of the Picture Coding Symposium}, San Jose,
  California, USA, December 2013, pp. 321--324.

\bibitem{kannala2006genericmodel}
J.~Kannala and S.~S. Brandt,
\newblock ``{A Generic Camera Model and Calibration Method for Conventional,
  Wide-Angle, and Fish-Eye Lenses},''
\newblock {\em IEEE Transactions on Pattern Analysis and Machine Intelligence},
  vol. 28, no. 8, pp. 1335--1340, August 2006.

\bibitem{compvis2}
S.~Zingg, D.~Scaramuzza, S.~Weiss, and R.~Siegwart,
\newblock ``{MAV Navigation through Indoor Corridors Using Optical Flow},''
\newblock in {\em Proceedings of the IEEE International Conference on Robotics
  and Automation}, Anchorage, Alaska, USA, May 2010, pp. 3361--3368.

\bibitem{compvis}
G.~H. Lee, F.~Faundorfer, and M.~Pollefeys,
\newblock ``{Motion Estimation for Self-Driving Cars with a Generalized
  Camera},''
\newblock in {\em Proceedings of the IEEE Conference on Computer Vision and
  Pattern Recognition}, Portland, Oregon, USA, June 2013, pp. 2746--2753.

\bibitem{eichenseer2014dcorfish}
A.~Eichenseer and A.~Kaup,
\newblock ``{Coding of Dis\-tor\-tion-\-Corrected Fisheye Video Sequences Using
  H.265/HEVC},''
\newblock in {\em Proceesings of the IEEE International Conference on Image
  Processing}, Paris, France, October 2014, pp. 4132--4136.

\bibitem{blender}
``{Blender Version 2.71},'' http://www.blender.org.

\bibitem{blendswap}
``{Blend Swap},'' http://www.blendswap.com.

\end{thebibliography}

\end{document}